\documentclass[useAMS,usenatbib]{mn2e} 
\usepackage{graphicx}
\usepackage{epsfig}
\usepackage{amsmath}

\newcommand{\hs}{\hspace{1mm}} 
\newcommand{\apj}{ApJ}
\newcommand{\aap}{A\&A} 
\newcommand{\apjl}{ApJL}
\newcommand{\mnras}{MNRAS} 

\newcommand{\apjs}{ApJS} 
\newcommand{\nat}{{\it Nature}}
\newcommand{\physrep}{PhR}
 
\newcommand{\pasj}{PASJ}

\newcommand\ion[2]{#1$\;${\scshape{#2}}}%                       % ion, i.e., CII = \ion{C}{ii}

\def\lsim{~\rlap{$<$}{\lower 1.0ex\hbox{$\sim$}}}

\def\gsim{~\rlap{$>$}{\lower 1.0ex\hbox{$\sim$}}}

\title[Ly$\alpha$ From the First Galaxies]{Seeing Through the Trough: Outflows and the Detectability of Ly$\alpha$ Emission from the First Galaxies}

\author[Dijkstra \& Wyithe]{Mark
Dijkstra$^{1}$\thanks{E-mail:mdijkstr@cfa.harvard.edu} and J. Stuart
B. Wyithe$^{2}$\\ $^{1}$Astronomy Department, Harvard University, 60
Garden Street, Cambridge, MA 02138, USA\\ $^{2}$School of Physics,
University of Melbourne, Parkville, Victoria, 3010, Australia}

\begin{document}

\date{\today} \pagerange{\pageref{firstpage}--\pageref{lastpage}}
\pubyear{2009}

\maketitle

\label{firstpage}
\begin{abstract}
The next generation of telescopes aim to directly observe the first
generation of galaxies that initiated the reionization process in our
Universe. The Ly$\alpha$ emission line is robustly predicted to be the
most prominent intrinsic spectral feature of these galaxies, making it
an ideal target to search for and study high redshift
galaxies. Unfortunately the large Gunn-Peterson optical depth of the
surrounding neutral intergalactic medium (IGM) is thought to render
this line extremely difficult to detect prior to reionization. In this
paper we demonstrate that the radiative transfer effects in the
interstellar medium (ISM), which cause Ly$\alpha$ flux to emerge from
galaxies at frequencies where the Gunn-Peterson optical
depth is reduced, can substantially enhance the prospects for
detection of the Ly$\alpha$ line at high redshift. In particular,
scattering off outflows of interstellar \ion{H}{I} gas can modify the
Ly$\alpha$ spectral line shape such that $\gsim 5\%$ of the emitted
Ly$\alpha$ radiation is transmitted directly to the observer, {\it
even through a fully neutral IGM}. It may therefore be possible to
directly observe `strong' Ly$\alpha$ emission lines (EW$\gsim 50$
\AA\hs rest frame) from the highest redshift galaxies that reside in
the smallest \ion{H}{II} `bubbles' early in the reionization era with
JWST. In addition, we show that outflows can boost the fraction
of Ly$\alpha$ radiation that is transmitted through the IGM during the
latter stages of reionization, and even post-reionization. Coupled
with the fact that the first generation of galaxies are thought to
have very large intrinsic equivalent Ly$\alpha$ equivalent widths,
 our results suggest that the search for galaxies in their
redshifted Ly$\alpha$ emission line can be competitive with the
drop-out technique out to the highest redshifts that can be probed 
in the JWST era.
  \end{abstract}

\begin{keywords}
galaxies: high redshift -- (galaxies): intergalactic medium -- ISM: kinematics and dynamics -- line: profiles -- scattering -- radiative transfer
\end{keywords}
 
\section{Introduction}
\label{sec:intro}

One of the main science drivers of the next generation of telescopes
is to detect the first generation of galaxies\footnote{The definition
of `the first generation of galaxies' is somewhat arbitrary. We take
it to mean any star forming galaxy during the earliest stages of the
reionization that might be detected (in  either their continuum or in
on of their lines) by the next generation of telescopes. Although the
analysis presented in this paper also applies to the first stars that
likely formed one--by--one in minihalos, the overall fluxes from these
sources is much too faint to be detected in the near future
(regardless of radiative transfer effects that may boost the
detectability of Ly$\alpha$ emission from such sources).} that formed
in our Universe. These galaxies contained hotter and more compact
stars \citep[e.g.][]{TS00,Bromm01}, whose initial mass function (IMF)
was likely top-heavy \citep[][]{La98,Bromm02}. Both the top-heavy
IMF and low (or zero) gas metallicity enhanced the number of
ionizing photons that the first galaxies emitted compared to that of
local galaxies, at a fixed star formation rate
\citep{TS00,Bromm01,S02,S03}. This enhancement in ionizing luminosity
results in larger \ion{H}{II} regions in the interstellar medium (ISM)
of these galaxies. As a result, one of the key predicted properties of
the first galaxies are prominent nebular emission lines, dominated in
flux by hydrogen Ly$\alpha$ ($\lambda=1216$\hs\AA, see e.g. Johnson et
al. 2009). The first generation of galaxies are therefore likely to have been
strong Ly$\alpha$ emitters, with equivalent widths possibly as high as
EW$\sim 1500$ \AA \hs\citep[][also see Partridge \& Peebles 1967,
Meier 1976]{S02,S03,J09b}.

In this paper we investigate the prospects for detecting this
Ly$\alpha$ emission. The first galaxies were surrounded by a mostly
neutral intergalactic medium (IGM), which is extremely optically thick
to Ly$\alpha$ radiation.
 \citet{LR99} showed that scattering of Ly$\alpha$ photons through a
 neutral IGM causes galaxies to be surrounded by diffuse
 Ly$\alpha$ halos \citep[also see][]{Ko04,Ko06}. While the total flux
 in these halos can be substantial, their large angular size results
 in surface brightness levels that are beyond reach of even future
 telescopes such as the {\it James Webb Space Telescope} (JWST,
 see \S~\ref{sec:RL}).

If Ly$\alpha$ radiation from the first galaxies is to be detected, it
will therefore be via Ly$\alpha$ photons that were transmitted
directly to the observer. In a neutral IGM, there are two mechanisms
by which this can be achieved:

\noindent ({\it i}) The first mechanism is due to ionizing radiation
that escapes from the galaxy, which creates a surrounding \ion{H}{II}
`bubble', that can strongly boost the detectability of Ly$\alpha$
emission\footnote{Indeed, the fact that the reionization process
likely affects the observed number and distribution of high--redshift
Ly$\alpha$ emitting galaxies is exactly the reason why Ly$\alpha$
emitting galaxies are thought to probe this epoch
\citep[][]{HS99,MR04,F06,Ka06,LF,McQ,Iliev08,Mesinger,Dayal10}.}
\citep[e.g.][]{Haiman02,Cen05}. This is because Hubble expansion
redshifts Ly$\alpha$ photons while they propagate freely through the
ionized gas. As as result, a fraction of photons enter the neutral
intergalactic medium on the red side of the line center, where the IGM
optical depth optical depth can be smaller by orders of magnitude. For
example, the Gunn-Peterson optical depth at redshift $z$ is given
by \citep[e.g.][]{BL01}

\begin{equation}
\tau_{\rm GP,0}\approx 7.30\times 10^5 x_{\rm HI}\Big{(}
\frac{1+z}{10}\Big{)}^{3/2},
\label{eq:taugp}
\end{equation} where $x_{\rm HI}$ denotes the neutral volume fraction of hydrogen in the IGM. The Gunn-Peterson optical depth (Eq~\ref{eq:taugp}) reduces to \citep{M98,DW06}
\begin{equation}
\tau_{\rm GP}(\Delta v)\approx 2.3\Big{(} \frac{\Delta v}{600\hs{\rm
km\hs s}^{-1}}\Big{)}^{-1}\Big{(} \frac{1+z}{10}\Big{)}^{3/2}
\label{eq:redgp}
\end{equation} for photons that enter the neutral IGM with a redshift of $\Delta v$  from line center. Eq~\ref{eq:redgp} illustrates that photons that enter the neutral IGM significantly redward of the Ly$\alpha$ resonance are scattered with only a weak optical depth. However given the high density of the surrounding IGM it is generally assumed that the first galaxies would not have had large enough \ion{H}{II} `bubbles' to prevent complete damping of the line.

\noindent ({\it ii}) The second mechanism for enhancing the
transmission of Ly$\alpha$ photons from high redshift galaxies is due
to  radiative transfer effects within the ISM of galaxies (in
particular outflows of interstellar \ion{H}{I} gas), which can shift
Ly$\alpha$ photons to the red side of the line before it reaches the
IGM. Observed interstellar metal absorption lines (\ion{Si}{II},
\ion{O}{I}, \ion{C}{II}, \ion{Fe}{II} and \ion{Al}{II}) in Lyman Break
Galaxies (LBGs) are typically strongly redshifted relative to the
galaxies' systemic velocity (with a median off-set of $\sim 160$ km
s$^{-1}$), while the Ly$\alpha$ emission line is strongly redshifted
\citep[with a median velocity offset of $\sim 450$ km s$^{-1}$ ][also
see Shapley et al. 2003]{Steidel10}. This suggest that large scale
outflows are ubiquitous in LBGs
\citep{Shapley03,Steidel10}. Furthermore, scattering of Ly$\alpha$
photons by \ion{H}{I} in outflows can successfully explain observed
Ly$\alpha$ line shapes in Ly$\alpha$ emitting galaxies at $z=3-6$
\citep[][]{V06,V08,V10}.

In this paper we explore the outflow mechanism in more detail. We will
show that these `local' (i.e. inherent to the galaxy itself) processes
in the ISM can cause as much as $\gsim 5\%$ of the emitted Ly$\alpha$
radiation to be directly transmitted to the observer {\it even through
a fully neutral IGM}. This result is important for the study of high redshift galaxies, because it
suggest that detecting the Ly$\alpha$ emission line of the first
generation of galaxies may well be within reach of the next generation
of telescopes including JWST.

The outline of this paper is as follows: we describe our models and
present our results in \S~\ref{sec:results}. We discuss our models and the
implications of this work in \S~\ref{sec:disc}, before we conclude in \S~\ref{sec:conc}. The
cosmological parameter values used throughout our discussion are
$(\Omega_m,\Omega_{\Lambda},\Omega_b,h)=(0.27,0.73,0.046,0.70)$
\citep{Komatsu08}.
\begin{figure*}
\vbox{\centerline{\epsfig{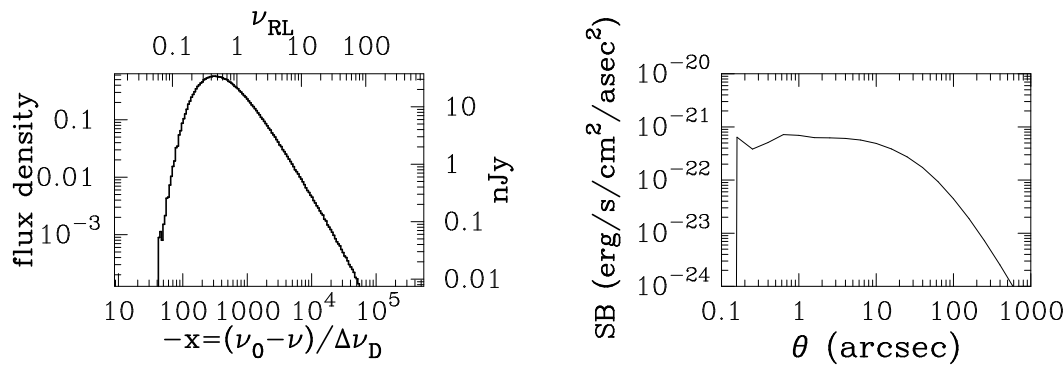}}}
\caption[]{The observed properties of Ly$\alpha$ halos (a.k.a
Loeb-Rybicki halos) that surround galaxies embedded within a fully
neutral comoving IGM. The `noisy' features in these plots are due to the finite (here
$N_{\rm phot}=10^6$) number of photons used in our Monte-Carlo
calculations. {\it Left panel:} the {\it histogram} shows the
integrated (over the entire area on the sky) emerging spectrum, under
the assumption that all Ly$\alpha$ photons were emitted at line
center, and assuming an emitted Ly$\alpha$ luminosity of
$L_{\alpha}=10^{43}$ erg s$^{-1}$. The lower horizontal axis shows the
dimensionless frequency parameter $x$, while the upper horizontal axis
shows the frequency parameter that was employed by Loeb \& Rybicki
(1999). The peak flux density occurs at $x\sim -300$ which in our
model corresponds to a redshift of $\sim 660$ km s$^{-1}$. The FWHM
of the spectrum is $\sim 1500$ km s$^{-1}$. The vertical axes are
expressed in arbitrary units on the left, and physical units on the
right. The peak integrated flux density reaches 30 nJy. For comparison
NIRSpec aboard JWST is expected to reach a sensitivity of $\sim$
hundreds of nJy in $10^4$ s (S/N=10) for R=2700, or line fluxes of
$\gsim 10^{-19} $ erg s$^{-1}$ cm$^{-2}$. {\it Right panel:} the
integrated (over frequency) surface brightness $S$ as a function of
impact parameter ($\theta$) from the galaxy. We find that $S<
10^{-21}$ erg s$^{-1}$ cm$^{-2}$ at all $\theta$. This plot illustrates that
it will be extremely difficult to directly detect these Ly$\alpha$
halos (see text). }
\label{fig:rl}
\end{figure*}

\section{Modelling of Ly$\alpha$ from high redshift galaxies}

We divide our study into modelling of three phenomena which are
important in the apparent brightness of Ly$\alpha$ emitters at high
redshift. Firstly in \S~\ref{sec:RL} we compute the integrated (over
the sky) flux density, and integrated (over frequency) surface
brightness profiles of the Loeb-Rybicki halos that surround the first
first generation of galaxies. This calculation illustrates the
difficulty of directly detecting these Ly$\alpha$ halos. Then in
\S~\ref{sec:static} we show how radiative transfer through a static
ISM causes the predicted Loeb-Rybicki halos to be accompanied by
Ly$\alpha$ point sources, which are more easily detectable. Finally in
\S~\ref{sec:outflow} we show that this result is strengthened further
when the ISM of galaxies contain outflows of \ion{H}{I}. In all our
calculations we assume that the intrinsic, i.e. the emitted,
luminosity of the galaxy is $L_{\alpha}=10^{43}$ erg s$^{-1}$, which corresponds to a star formation
rate of SFR$\lsim 2 M_{\odot}$ yr$^{-1}$ for $Z\lsim 10^{-3}Z_{\odot}$
\citep[][for a Salpeter IMF from $1-100M_{\odot}$]{S03}. For
comparison, the brightest observed Ly$\alpha$ sources at $z=6.5-6.6$
have `observed' luminosities (i.e. the observed flux times $4\pi
d^2_L(z)$, where $d_L(z)$ denotes the luminosity distance to redshift
$z$) of a few times $10^{43}$ erg s$^{-1}$
\citep{Ka06,Ouchi10}. Depending on what fraction of the emitted
Ly$\alpha$ flux we detect, their intrinsic luminosities can be
significantly higher. Our results scale linearly with
$L_{\alpha}$.

Throughout this paper, we denote photon frequency with the
dimensionless parameter  $x=(\nu-\nu_{\alpha})/\Delta \nu_{\alpha}$,
in which $\nu_{\alpha}=2.47\times 10^{15}$ Hz denotes the Ly$\alpha$
restframe frequency; $\Delta \nu_{\alpha}=\nu_{\alpha}v_{\rm th}/c$,
$v_{\rm th}=\sqrt{2k_B T/m_p}=12.9(T/10^4\hs{\rm K})^{1/2}$ km
s$^{-1}$, where $T$ denotes the gas temperature. Here, $c$, $k_B$, and
$m_p$ denote the speed of light, Boltzmann constant, and the proton
mass, respectively. In all our calculations we assume that the gas
temperature in the ISM is $T_{\rm ISM}=10^4$ K, but we have explicitly
verified that our final results do not depend on this assumption.
Furthermore, the Gunn-Peterson optical depth for photons that enter
the IGM while in the red wing of the Ly$\alpha$ line is independent of
temperature (see Eq~\ref{eq:redgp}), and the assumed IGM temperature
of $T_{\rm IGM}=300$ K (appropriate for neutral intergalactic gas, see e.g. Pritchard \& Loeb 2008) is irrelevant.

\label{sec:results}
\subsection{Loeb-Rybicki Halos}
\label{sec:RL}

Figure~\ref{fig:rl} shows the observed properties of Ly$\alpha$ halos
that surround galaxies embedded within a fully neutral comoving IGM.
The {\it histogram} in the {\it left panel} shows the integrated (over
the entire area on the sky) emerging spectrum computed with the
Monte-Carlo Ly$\alpha$ radiative transfer code McHammer \citep{D06a}. Following previous work, we assumed in this calculation that the Ly$\alpha$
photons were emitted at line center, i.e. $x=0$. The lower horizontal
axis shows the dimensionless frequency parameter $x$, while the upper
horizontal axis shows the frequency parameter that was employed by
Loeb \& Rybicki (1999). All Ly$\alpha$ photons emerge with a systemic redshift (i.e. $x \ll 0$). The peak flux
density occurs at $x\sim -300$ which corresponds to a redshift of
$\sim 660$ km s$^{-1}$. The Full Width at Half Maximum (FWHM) of the
spectrum is $\sim 1500$ km s$^{-1}$. The vertical axes contain
arbitrary units on the left, and physical units on the right. The peak
integrated flux density is $\sim$30 nJy.

The {\it right panel} shows the integrated (over frequency) surface
brightness $S$ (in erg s$^{-1}$ cm$^{-2}$ arcsec$^{-2}$) as a function
of impact parameter ($\theta$, angular separation) from the galaxy in
arcsec. The surface brightness contains a core out to $\theta\sim 20$
arcsec after which it drops steadily. We find that $S< 10^{-21}$ erg
s$^{-1}$ cm$^{-2}$ at all $\theta$. For comparison, \citet{Rauch08}
reached a $1-\sigma$ surface brightness limit of $S_{\rm lim}=8 \times
10^{-20}$ erg s$^{-1}$ cm$^{-2}$ in a 92 hr long exposure with the ESO
VLT-FORS2 instrument, which represent the deepest observations to
date. This, when combined with JWST's proposed sensitivity
limit\footnote{See
http://www.stsci.edu/jwst/science/sensitivity/. This estimate assumes
$R=1000$ or $R=2700$, and assumes an integration time of $10^5$ s.} to
point sources of $\gsim 10^{-19} $ erg s$^{-1}$ cm$^{-2}$, implies
that it is and will remain extremely difficult to detect these
Ly$\alpha$ halos, even with the next generation of telescopes.

\subsection{Static ISM}
\label{sec:static}

In the previous calculation we assumed that all Ly$\alpha$ escaped
from the galaxy at the line center. In reality complex radiative transfer
effects occur inside the ISM of a galaxy. \ion{H}{I} observations show
that local galaxies contain \ion{H}{I} column densities of $N_{\rm
HI}\sim 10^{19}-10^{21}$ cm$^{-2}$. Such large columns of \ion{H}{I}
gas can affect the spectrum of Ly$\alpha$ photons as they emerge from
the galaxy. Furthermore, we expect column densities of
gas in high redshift galaxies to increase as  $\propto (1+z)^2$ (see
the end of this section for a more quantitative discussion). It is
therefore plausible that Ly$\alpha$ photons need to traverse a
substantial, i.e. $N_{\rm HI}> 10^{20}$ cm$^{-2}$, column of
\ion{H}{I} gas before escaping from the galaxy. To estimate the impact of large \ion{H}{I} column densities on the
detectability of Ly$\alpha$ emission, we turn to analytic solutions
for the spectrum of Ly$\alpha$ photons emerging from
static, homogeneous, extremely optically thick media. Following the analyses of \citet{Harrington73} and \citet{Neufeld90},
\citet{D06a} computed the Ly$\alpha$ spectrum emerging from a sphere
to be

\begin{equation}
J(x)=\frac{\pi^{0.5}}{\sqrt{24}a\tau_0}\Bigg{(}\frac{x^2}{1+{\rm
cosh}\Big{[}\sqrt{\frac{2\pi^3}{27}}\frac{|x^3|}{a\tau_0}\Big{]}}\Bigg{)},
\label{eq:slab}
\end{equation} assuming a source of photons in the center of the sphere. Here, $\tau_0=5.9\times 10^6(N_{\rm HI}/10^{20}\hs{\rm cm}^{-2})(T_{\rm ISM}/10^4\hs{\rm K})^{-1/2}$ denotes the line center optical depth from the center to the edge of the sphere, and $a=4.7\times 10^{-4}(T_{\rm ISM}/10^4\hs{\rm K})^{-1/2}$ denotes the Voigt parameter. The function $J(x)$ is normalized to $(2\pi)^{-1}$, and its maximum occurs at $x_{\rm p}\pm0.9(a_v\tau_0)^{1/3}\sim \pm 13 (N_{\rm HI}/10^{20}\hs{\rm cm}^{-2})^{1/3}(T_{\rm ISM}/10^4\hs{\rm K})^{-1/3}$. This dimensionless frequency shift translates to $\Delta v=\pm 160(N_{\rm HI}/10^{20}\hs{\rm cm}^{-2})^{1/3}(T_{\rm ISM}/10^4\hs{\rm K})^{1/6}$ km s$^{-1}$. The weak temperature dependence ($\propto T^{1/6}$) explains the weak dependence of our results on the assumed ISM temperature. The Ly$\alpha$ spectrum emerging from a semi-infinite slab with the same line center optical depth is broader by a factor of $1.1$, but is otherwise very similar. 

We compute how the predicted properties of the observed Ly$\alpha$
radiation change when we inject photons into a neutral IGM following
the distribution given by Eq~\ref{eq:slab}, rather than in the line
center (as in \S~\ref{sec:RL}).
\begin{figure}
\vbox{\centerline{\epsfig{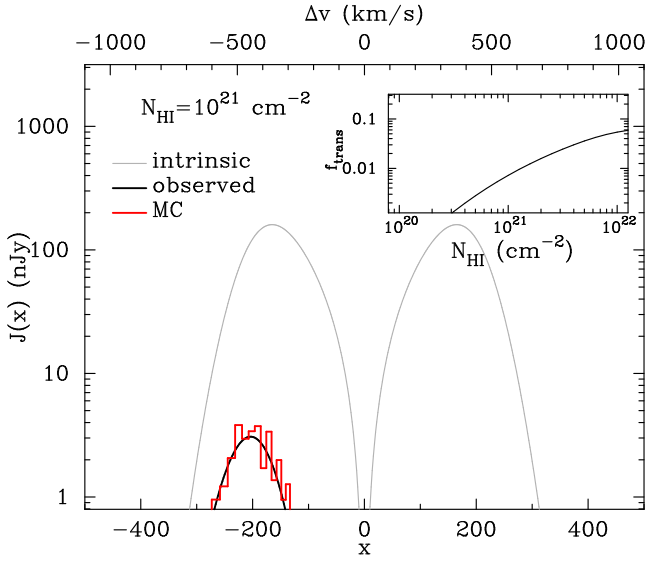}}}
\caption[]{The {\it solid histogram} in Figure shows the spectrum of
Ly$\alpha$ photons that were not scattered at all. These unscattered
photons would be observed as a point source. The {\it grey line}
shows the {\it intrinsic} flux density of the source, $J_{\rm
int}(x)$, which corresponds to the observed Ly$\alpha$ spectrum if we
were able to detect all Ly$\alpha$ radiation
(Eq~\ref{eq:slab}). Overplotted as the {\it black solid line} shows
the analytic calculation of the observed spectrum $J_{\rm point}(x)$
given by $J_{\rm int}(x)\exp(- \tau_{\rm IGM}[x])$ (see text). The
{\it inset} shows the fraction of directly transmitted Ly$\alpha$
flux, $f_{\rm trans}$, as a function of \ion{H}{I} column density
$N_{\rm HI}$. The transmitted fraction increases from $\sim 1\%$ at
$N_{\rm HI}=10^{21}$ cm$^{-2}$ to $\gsim 5\%$ at $N_{\rm HI}\gsim
7\times 10^{21}$ cm$^{-2}$. A transmission $f_{\rm trans}$ of only a
few per cent is important: if the intrinsic Ly$\alpha$ EW of the first
generation of galaxies was as high as $1500$ \AA, then the observed
restframe EW is EW$\sim 50(f_{\rm trans}/0.03)$ \AA (see text)}
\label{fig:static}
\end{figure}
The most important difference is that a non-zero fraction of the
photons are transmitted directly to the observer without
scattering\footnote{The scattered radiation is spread out in a halo
that resembles the Loeb-Rybicki halos. However, at small angular
separations ($\theta \lsim 1$ arcsec) we find that the surface
brightness profile is enhanced by a factor of $\sim$a few-ten (see
Appendix~\ref{app:halo} for an example). This is still well below
detection thresholds of existing and future facilities. Furthermore,
this enhancement in the central surface brightness profile vanishes
when we allow for the existence of an \ion{H}{II} region around the
galaxy (see Fig~\ref{fig:RLout}).}. The {\it solid histogram} in
Figure~\ref{fig:static} shows the spectrum of Ly$\alpha$ photons that
were not scattered at all. These unscattered photons would be observed
as a `point source' (the angular scale is set by the physical scale of
the scattering medium), and its peak flux density is $\sim 4$
nJy. This peak flux density is smaller than the peak integrated flux
density of the Loeb-Rybicki halos, because in this model the total
fraction the photons that was transmitted to the observer directly,
$f_{\rm trans}$, was only $\sim 1\%$. For comparison, we have
overplotted the {\it intrinsic} flux density of the source as the {\it
grey line}, as given by Eq~\ref{eq:slab}. This corresponds to the
observed Ly$\alpha$ spectrum if we were able to detect all Ly$\alpha$
radiation. Overplotted ({\it black solid line}) is the analytic
calculation of the observed spectrum $J_{\rm point}(x)$ given by

\begin{equation}
J_{\rm point}(x)=J(x)\exp(- \tau_{\rm IGM}[x]),
\label{eq:point}
%\label{eq:trans}
\end{equation} where $J(x)$ is given by Eq~\ref{eq:slab}. Furthermore, $\tau_{\rm IGM}(x)$ denotes the opacity of the neutral IGM to a Ly$\alpha$ photon that escapes from the galaxy at frequency $x$, which is $\tau_{\rm IGM}(x)=\tau_{\rm GP,0}\frac{1}{\sqrt{\pi}}\int_{-\infty}^{x}\phi(x')dx'$ \citep[e.g. Appendix A of][]{D06a}\footnote{This implicitly assumes that the gas in IGM is at mean density of the Universe, and that it is co-moving with the Hubble flow. In reality we expect strong departures from the Hubble flow in close proximity to halos, and that the gas is overdense there \citep{Infall}. If we incorporate this into our model, then this increases the opacity of the IGM. However, ionizing radiation that escapes from the galaxy ionizes most of this denser gas. Indeed, we have verified that for escape fractions of ionizing radiation as low as $f_{\rm esc}=1\%$, more realistic models of the IGM \citep[as in][]{IGM} actually transmit more Ly$\alpha$. The calculations that are present in this paper are therefore conservative.}. Here, $\tau_{\rm GP,0}$ denotes the Gunn-Peterson optical depth given by Eq~\ref{eq:taugp}, and $\phi(x)$ denotes the Voigt function \citep{RL79}. Not suprisingly, the analytic calculation agrees well with the Monte-Carlo calculation (which in the end should generate random realizations of this function). We point out that the predicted Ly$\alpha$ spectral line shape is not as asymmetric as is often observed in lower redhsift galaxies. This is because the Ly$\alpha$ absorption cross-section varies weakly with frequency (as $\Delta v^{-1}$, see Eq~\ref{eq:redgp}) in its damping wing, and scattering in the IGM introduces no 'sharp' cut-off in the observed Ly$\alpha$ line shape.

The {\it inset} of Figure~\ref{fig:static} shows how the fraction of
directly transmitted Ly$\alpha$ flux depends on \ion{H}{I} column
density $N_{\rm HI}$. As mentioned previously, $f_{\rm trans}\sim
10^{-2}$ at $N_{\rm HI}=10^{21}$ cm$^{-2}$. The transmitted fraction
increases to $\gsim 5\%$ at $N_{\rm HI}\gsim 7\times 10^{21}$
cm$^{-2}$. For comparison, the central column density for a standard
Mo-Mao \& White exponential disk--when seen face on--is $N_{\rm
HI,0}\sim 3\times 10^{22}([1+z]/11)^{3/2}(v_{\rm circ}/13 \hs {\rm
km}/{\rm s})$, under standard assumptions that the spin parameter
$\lambda=0.05$, and that $j_d=m_d=0.05$. This suggests that \ion{H}{I}
column densities in excess of $10^{21}$ cm$^{-2}$ are expected at
the redshifts of interest, and that therefore that $f_{\rm trans}$ can
exceed a few per cent.

{\it We stress that although small, a transmission $f_{\rm trans}$ of only a few per
cent is important}. The total observed flux in the Ly$\alpha$ point
source at $z=10$ is $f_{\rm obs}=3.7 \times 10^{-19}(f_{\rm
trans}/0.05)\sim 10^{-19}$ erg s$^{-1}$ cm$^{-2}$ which is within reach of JWST. For example, an $R=1000$
grating would yield a $5-\sigma(f_{\rm trans}/0.05)$ detection in
a $10^5$ s exposure (band I). In addition to rendering the
line emission from very high redshift galaxies detectable by JWST, this effect could lead to strong Ly$\alpha$, since given an intrinsic Ly$\alpha$ EW for the first generation of galaxies of EW$_{\rm
int}=1500$ \AA, the observed restframe EW is EW$\sim 50(f_{\rm
trans}/0.03)$ \AA. For comparison, only$\gsim 4 \%$ of LBGs at $z=3$
have emission lines with EW\gsim 50\AA\hs\citep{Shapley03}. Indeed, for an observed rest frame EW$\sim 75(f_{\rm
trans}/0.05)$ \AA, and a line flux of $f_{\rm obs}=3.7 \times
10^{-19}(f_{\rm trans}/0.05)\sim 10^{-19}$ erg s$^{-1}$ cm$^{-2}$ (typical for lower redshift Ly$\alpha$ emitters), the
corresponding continuum flux density is $f_{\nu}\sim 2.7$ nJy, which
would only yield a $\sim 2-\sigma$ detection in $10^5$ s with {\it
NIRCAM} on JWST. The Ly$\alpha$ line flux may therefore be
(slightly) more easily detectable than the continuum emission, even
when we only observe a few per cent of the emitted Ly$\alpha$ flux.

\begin{figure}
\vbox{\centerline{\epsfig{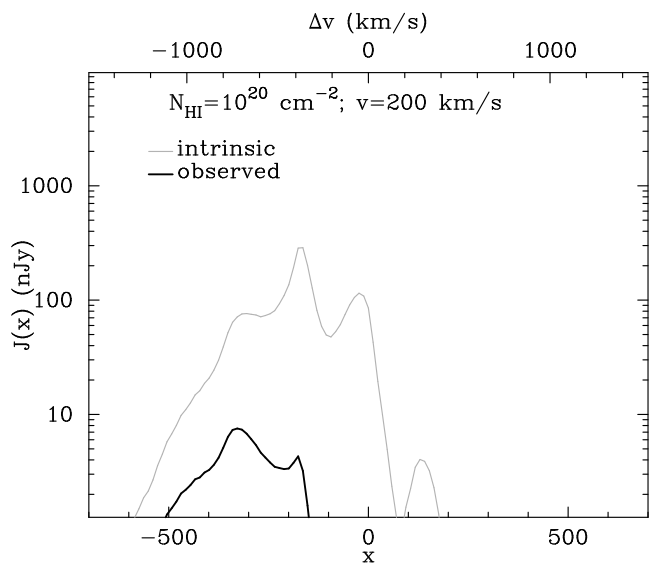}}}
\caption[]{The {\it black solid line} shows the spectrum of Ly$\alpha$
photons that were transmitted directly to an observer through a fully
neutral IGM. The {\it grey solid line} shows the intrinsic spectrum of
this galaxy (see Ahn et al. 2003 and Verhamme et al. 2006 for a
detailed discussion on these features in the spectrum). The ISM of
this galaxy was modeled as a thin, outflowing (with speed $v_{\rm
sh}$), spherically symmetric shell of \ion{H}{I} gas (with column
density $N_{\rm HI}$). This type of model successfully reproduces
observed Ly$\alpha$ line shapes in known Ly$\alpha$ emitting
galaxies. We assumed that  $(N_{\rm HI},v_{\rm sh})=(10^{20}\hs{\rm
cm}^{-2}, 200\hs{\rm km} \hs {\rm s}^{-1})$. Because a significant
fraction of the radiation comes out of the galaxy with a large
systematic redshift, $f_{\rm trans}\sim 4\%$  of all emitted photons
is transmitted directly to the observer.}
\label{fig:out}
\end{figure}

\begin{figure}
\vbox{\centerline{\epsfig{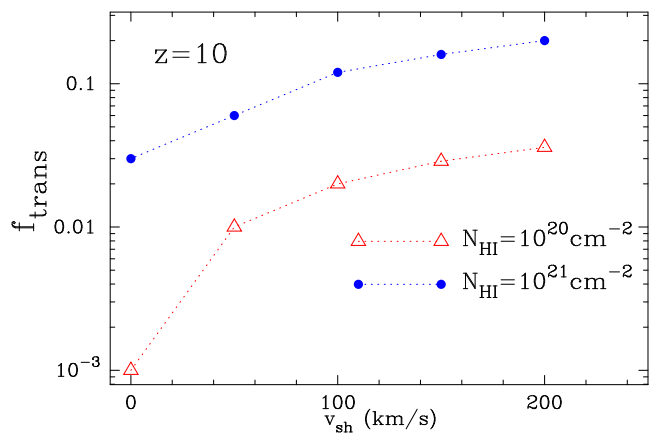}}}
\caption[]{This figure shows the fraction of Ly$\alpha$ photons,
$f_{\rm trans}$, that is transmitted directly through a fully neutral
IGM at $z=10$, as a function of $v_{\rm sh}$ for $N_{\rm HI}=10^{20}$
cm$^{-2}$ ({\it red squares}) and $N_{\rm HI}=10^{21}$ cm$^{-2}$ ({\it
blue squares}). Outflows of $\gsim 50 $ km s$^{-1}$ can boost $f_{\rm
trans}$ tremendously to values as large as $f_{\rm trans}\sim 10-20\%$
(for $v_{\rm sh}=200$ km s$^{-1}$). That is, $10-20\%$ of all
Ly$\alpha$ photons may be transmitted through a fully neutral IGM.}
\label{fig:plot}
\end{figure}
\begin{figure}
\vbox{\centerline{\epsfig{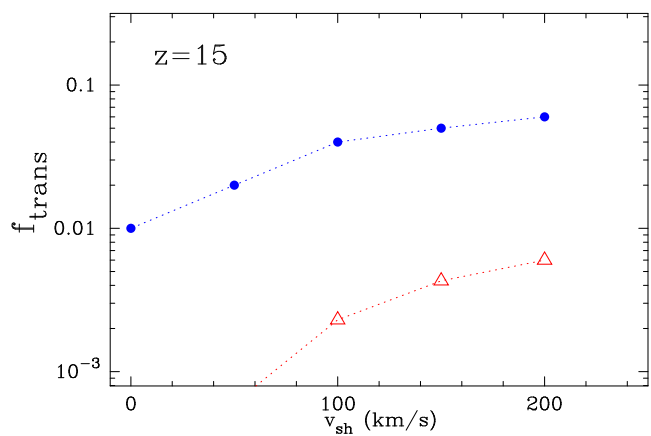}}}
\caption[]{Same as Figure~\ref{fig:plot}, but at $z=15$.}
\label{fig:plot2}
\end{figure}

\subsection{Outflowing ISM}
\label{sec:outflow}

Observed galaxies typically show evidence that large scale outflows
are ubiquitous in LBGs: interstellar metal absorption lines
(\ion{Si}{II}, \ion{O}{I}, \ion{C}{II}, \ion{Fe}{II} and \ion{Al}{II})
in Lyman Break Galaxies (LBGs) are typically strongly redshifted
relative to the galaxies' systemic velocity (with a median off-set
of $\sim 160$ km s$^{-1}$), while the Ly$\alpha$ emission line is
strongly redshifted \citep[with a median velocity offset of $\sim 450$
km s$^{-1}$ ][also see Shapley et al. 2003]{Steidel10}. Scattering of
Ly$\alpha$ photons by \ion{H}{I} in these outflows has successfully
explained observed Ly$\alpha$ line shapes--and their redshifts--in
Ly$\alpha$ emitting galaxies at $z=3-6$ \citep[][see
\S~\ref{sec:model} for a discussion on this model]{V06,V08,V10}.

We therefore repeat the exercise of \S~\ref{sec:static} for a suite of
outflow models. Following \citet{V06,V08}, we model the outflow as a
spherically symmetric thin shell of gas that contains an \ion{H}{I}
column density $N_{\rm HI}$ \citep[also see][]{Ahn03}, and outflow
velocity $v_{\rm sh}$. We assume that the shell has a radius of $1$
kpc and a thickness of 0.1 kpc, but stress that the precise physical
scale of the outflow is not important for our results. Our assumed gas
temperature in the outflowing \ion{H}{I} shell of $T_{\rm ISM}=10^4$ K
corresponds to a $b$-parameter of $b\sim 13$ km s$^{-1}$ in the
terminology of Verhamme et al. (2008). Our results do not depend on
this choice for $b$, as the amount of flux at large $\Delta v$ depends
very weakly on this parameter (see the {\it right panel} of Fig~15 of
Verhamme et al. 2006). We further assume the \ion{H}{I} shells to be
dust-free (see \S~\ref{sec:dust}). \citet{V08} typically found that
$\log N_{\rm HI}\sim 19-22$, and $v_{\rm sh}\sim 0-500$ km s$^{-1}$,
and this is the range of parameter space we explore. We compute
Ly$\alpha$ spectra emerging from the outflows with the Monte-Carlo
transfer code \citep{D06a}. In our calculations, the Ly$\alpha$
photons are emitted at line center, i.e $x=0$. We verified that our
results are insensitive to this assumption\footnote{We repeated some
calculations in which Ly$\alpha$ photons were emitted following a
Gaussian distribution with a standard deviation of $\sigma=100$ km
s$^{-1}$. While the precise emerging line shapes changed, the overall
transmitted fraction remained practically identical.}. We compute the
impact of the IGM on the directly observed fraction of Ly$\alpha$ by
simply suppressing the intrinsic spectrum by exp$(- \tau_{\rm
IGM}[x])$ (see \S~\ref{sec:static}).

The {\it grey solid line} in Figure~\ref{fig:out} shows an example of
the intrinsic spectrum for a model in which we assumed that $(N_{\rm
HI},v_{\rm sh})=(10^{20}\hs{\rm cm}^{-2},200\hs{\rm km} \hs {\rm
s}^{-1})$. The intrinsic spectrum is highly asymmetric, with more flux
coming out on the red side of the Ly$\alpha$ line center. The spectrum
peaks at about $\sim 2v_{\rm sh}$, as expected for radiation that
scatters back to the observer on the far side of the galaxy (see
Verhamme et al. 2006, and Ahn et al. 2003 for a detailed
discussion on these features in the spectrum). However, a significant
fraction of the radiation comes out at larger redshifts. This
radiation can be transmitted directly to an observer through a neutral
IGM, and would be observed as a point-source. The {\it
black solid line} shows the spectrum of the point source (see
Appendix~\ref{app:halo} for a plot of the surface brightness profile
of the scattered radiation). We find that the spectrum of the point
source contains $f_{\rm trans}\sim 4\%$  of all emitted photons. Note
that the FWHM of the observed spectrum of the galaxy, as well as the
offset arise purely from radiative transfer effects.

The directly transmitted fraction $f_{\rm trans}$ depends on both
$N_{\rm HI}$ and $v_{\rm sh}$, In Figure~\ref{fig:plot} we show
$f_{\rm trans}$ as a function of $v_{\rm sh}$ for $N_{\rm HI}=10^{20}$
cm$^{-2}$ ({\it red squares}) and $N_{\rm HI}=10^{21}$ cm$^{-2}$ ({\it
blue squares}). Having outflows as low as $\gsim 50 $ km s$^{-1}$
makes a significant difference in the predicted  $f_{\rm trans}$. For
example, we find $f_{\rm trans}\gsim 5\%$ for $v_{\rm sh}\gsim 200$ (50)
km s$^{-1}$ for log$N_{\rm HI}=20$ (log$N_{\rm HI}=21$). We stress
that the outflow parameters were chosen to lie the range required to
explain line shapes of observed LAEs at $z<6$. If the dark
matter halos that host the highest redshift galaxies had lower masses
(as expected in a hierarchical structure formation scenario), then
outflows at higher redshift may have occured at lower speeds, because
of the observed (at lower redshifts) scaling of outflow velocity with
circular velocity of the host dark matter halo
\citep[e.g.][]{Martin05}. On the other hand, higher redshift galaxies
are expected to be more compact, and therefore we naively expect
larger column densities of \ion{H}{I} gas at higher redshifts (see
\S~\ref{sec:static}), in which case low outflow velocities can
boost $f_{\rm trans}$ tremendously. The outflow properties in the
highest redshift galaxies, and their impact on Ly$\alpha$ propagation
is clearly a topic that needs further investigation.

\section{Discussion}
\label{sec:disc}

In this section we discuss a range of issues arising from our model.

\subsection{Detecting Ly$\alpha$ Emission During and After the Epoch of Reionization}
\label{sec:EoR}

The results of the previous section focused on the first generation of
galaxies that were surrounded by a neutral IGM. However our work
applies more broadly. For example, radiative transfer effects in the
ISM can also boost the detectability of the Ly$\alpha$ line emitted by
galaxies during the later stages, and even post-reionization. It is
generally thought that ionized intergalactic gas is transparent to
Ly$\alpha$ radiation. More specifically, residual \ion{H}{I} that
exists inside the ionized IGM is assumed to only affect radiation
frequencies that lie blueward of the Ly$\alpha$ resonance. However, in
the standard cosmological model (ionized) intergalactic gas in close
proximity to galaxies is expected to be overdense ($1+\delta \sim
2-20$, Barkana 2004). Furthermore, gravity causes this gas to flow
towards galaxies. This inflowing, denser ionized gas is expected to
scatter a significant fraction of the Ly$\alpha$ photons out of the
line of sight. Even the ionized IGM therefore transmits on average as
little as $\sim 10-30 \%$ of the Ly$\alpha$ to an observer
\citep{IGM,Iliev08,ZZ,Dayal10}. It is therefore unclear whether
\ion{H}{II} regions during the reionization process actually transmit
enough Ly$\alpha$ flux to an observer, to render it detectable.

However in the presence of \ion{H}{I} outflows, the transmitted
fraction of Ly$\alpha$ radiation through ionized gas can increase
dramatically. This is because the outflowing \ion{H}{I} gas imparts
a redshift to the Ly$\alpha$ photons that is well in excess of $\sim$ a few
hundred km s$^{-1}$ (see \S~\ref{sec:outflow} and
Fig~\ref{fig:out}). On the other hand, the influence of the ionized
IGM only extends out to $\Delta v \lsim v_{\rm infall}$
\citep{Infall}. Here, the inflow velocity, $v_{\rm infall}$, is of
order the circular velocity of the dark matter halo hosting the
galaxy, $v_{\rm infall}\sim v_{\rm circ}=80(M_{\rm
halo}/10^{10}M_{\odot})^{1/3}([1+z]/11)^{1/2}$ km s$^{-1}$
\citep[e.g.][where $M_{\rm halo}$ denotes the dark matter halo
mass]{BL01}. The impact of the ionized IGM in close proximity to the
galaxy is therefore significantly weaker when \ion{H}{I} outflows are
important in shaping the Ly$\alpha$ spectral line shape that emerges
from galaxies.

This implies that winds can have important consequences when
interpreting the observed sudden drop in the Ly$\alpha$ luminosity
between $z=6.5$ and $z=5.7$ \citep[][also see Ouchi et al. in
prep]{Shima06,Ka06,Ota08}. The most important aspect of this observation is
that the rest-frame UV luminosity function of these same galaxies does
not evolve between these redshifts (within the uncertainties,
Kashikawa et al. 2006). Indeed, \citet{LF} have shown that these two
observations combined translate to a reduction in the number of
detected Ly$\alpha$ photons from $z=6.5$ by a factor of $1.1-1.8$
(95\% CL) relative to $z=5.7$. The simplest interpretation of this
observation is that the IGM at $z=6.5$ is more opaque to Ly$\alpha$
photons than at $z=5.7$, because the IGM naturally only affects the
observed Ly$\alpha$ flux. \citet{LF} argued that this can be explained
quite naturally by an evolution in the opacity of the ionized
IGM. However, if winds reduce the impact of the ionized component,
then this conclusion becomes uncertain, and the observed reduction in
the Ly$\alpha$ luminosity function may be at least partly a
reionization--induced signature. In order to understand the evolution
of the transmission in terms of the reionization history it is
therefore crucial to understand the properties of winds in
high--redshift galaxies, and how they affect the transport of
Ly$\alpha$ radiation.

\subsection{Detecting Other Lines}
\label{sec:lines}

In this paper we have focused our attention on the Ly$\alpha$ emission
line, because it is predicted to be the strongest spectral line of
high redshift galaxies. It may be possible to detect other spectral
lines as well.

The most prominent alternative emission line is H$\alpha$
($\lambda=6536$ \AA), which is intrinsically weaker than Ly$\alpha$ in
flux by a factor of $\sim 8$, under the assumption of case-B
recombination. However only a fraction $f_{\rm trans}=0.05$ of the
emitted Ly$\alpha$ radiation is transmitted, so that the observed
H$\alpha$ flux may be larger by a factor of $\sim 2.5(f_{\rm
trans}/0.05)^{-1}$. On the other hand the H$\alpha$ line lies deeper
in the IR, $\lambda= 7([1+z]/11)$ $\mu$m, where JWST's {\it Mid
Infrared Instrument} would yield a lower S/N detection of the
H$\alpha$ line for medium resolution spectroscopy. This suggests that
it may be more difficult to detect H$\alpha$ than naively expected.

In addition the Helium Balmer $\alpha$ (HeII $\lambda=$1640 \AA) line
could contain a flux that is comparable to that of H$\alpha$
\citep{Oh01,J09b}. However, the predicted strength of this line
depends sensitively on the IMF, and it can be weaker than H$\alpha$ by
an order of magnitude for reasonable model assumptions
\citep[][]{J09b}. On the other hand, the composite spectrum of {\it
observed} $z=3$ LBGs contains a HeII $\lambda=$1640 \AA\hs emission
line \citep{Shapley03}. The Full Width at Half Maximum of this line is
FWHM$\sim 1500$ km s$^{-1}$, which is broader than the FWHM of most
other nebular emission lines. The origin of this line is not
resolved. The line has been associated with population III star
formation \citep{JH06} occurring in pockets of pristine gas that
persisted down to low redshift, but may also originate in winds
associated with massive Wolf-Rayet stars. In this latter case the line
is expected to vanish for low gas metallicities \citep{Br}. Thus it is
not clear whether the detection of HeII $\lambda=$1640 \AA\hs in z=3
LBGs implies that this line should be present in the higher redshift
galaxies.

Finally, we expect numerous recombination lines associated with metals
heavier than Helium (such as [\ion{O}{II}], [\ion{O}{III}],
[\ion{S}{II}], ..., e.g. Huchra 1977). Again, robust predictions of
the strengths of these lines (which fall in the restframe optical) do
not exist as it requires an accurate knowledge of the metallicity of \ion{H}{II} regions in high--redshift galaxies.

\subsection{Dust}
\label{sec:dust}

Our calculations have ignored dust. To understand the impact of dust
on the Ly$\alpha$ radiation field requires understanding the
Ly$\alpha$ transfer process through the ISM of
galaxies\footnote{Several groups have included dust when modeling the
observed number density of Ly$\alpha$ emitters
\citep[e.g.][]{HS99,Mao07,Ko10,Dayal10}. When comparing such models
with actual data, a degeneracy exists between the opacity of the
ionized IGM and dust to Ly$\alpha$ radiation (i.e. predicted number
densities are not sensitive to what the exact source of opacity
is). As argued in this section, understanding how outflows shape the
intrinsic Ly$\alpha$ emission line that emerges from galaxies provide
such models with important new constraints on the IGM--and
therefore--dust opacity.}. This is because the scattering process
causes Ly$\alpha$ photons to traverse different paths through the ISM
than restframe UV continuum photons
\citep[][]{CF91,Neufeld91,Hansen06,Laursen09}. As a result dust has a
different impact on Ly$\alpha$ line and UV continuum
photons. Observations indicate the Ly$\alpha$ escape fraction
decreases with the dust content of galaxies. More specifically, the
mean Ly$\alpha$ equivalent width decreases with the observed reddening
of Lyman Break Galaxies \citep{Shapley03}. \citet{V08} have used their
models to quantify the actual escape fraction of Ly$\alpha$ photons
from galaxies, $f_{\rm d,esc}$, as a function of the observed dust
reddening, and found that on average $f_{\rm d,esc}\sim
10^{-7.7E(B-V)}$. \citet{Bouwens10} recently argued that candidate
galaxies at $z=7$ are considerably bluer than at lower
redshifts. Their most luminous candidates are consistent with
E(B-V)$\sim 0.05$, which translates to $f_{\rm d,esc}\sim 0.4$ (which
is significantly higher than Ly$\alpha$ escape fraction inferred for
galaxies at $z \lsim 2$ see e.g. Hayes et al. 2007,2010). If we take
this number literally, then our quoted Ly$\alpha$ fluxes are high by a
factor of $\sim 2.5$ (which implies that dust is not the dominant
uncertainty). Moreover, in hierarchical models of galaxy formation,
we expect galaxies at higher redshifts, and lower luminosities to be
less evolved, and hence the impact of dust is likely to be weaker.

The relation derived by \citet{V08} was obtained by averaging over a
dozen galaxies, and there exists considerable scatter around their
derived relation. For example, observations of low-redshift star
forming galaxies with the {\it International Ultraviolet Explorer}
(IUE) and the {\it Hubble Space Telescope} (HST) revealed very weak
Ly$\alpha$ emission (or even strong absorption) in metal poor objects
\citep{Hartmann84,Hartmann88,Kunth94}, while strong Ly$\alpha$
emission could be detected from some metal and dust rich galaxies
\citep[e.g.][]{L95}. This scatter is likely related to the precise
geometry of the dust distribution \citep[][]{Scarlata09}, viewing
angle towards the galaxies \citep{Laursen09}, and outflows. When
outflows are present in galaxies, the strength of Ly$\alpha$ emission
appears to be independent of the dust content of the galaxy. In the
absence of strong outflows however, Ly$\alpha$ line strength appears
to decrease with dust content \citep[][]{Kunth98,Atek08}. Our work
implies that in high redshift galaxies, \ion{H}{I} outflows further
affect the subsequent impact of the IGM on the detectability of
Ly$\alpha$ emitting galaxies. 

\subsection{Peculiar Velocities}
Following the announcement of the discovery of a $z=10$ galaxy by
\citet{Pello04}--which was later disputed \citep{Bremer04,W04}--,
\citet{Cen05} argued that the line could be detected through a fully
neutral IGM, if the Ly$\alpha$ emitting region in the galaxy was
receding relative to the surrounding absorbing IGM with $v\gsim 35$ km
s$^{-1}$. In this study, the galaxy ionized an \ion{H}{II} region with
a radius of $\sim 0.28$ pMpc. While peculiar velocities of this
magnitude do not play an important role in reducing the Gunn-Peterson
damping wing optical depth (for which one needs $\Delta v \gsim 500$
km s$^{-1}$, Eq~\ref{eq:redgp}), they can play a role in reducing the
opacity of the ionized `local' IGM (see \S~\ref{sec:EoR}), just like
outflows. However, both mechanisms are expected to leave different
signatures on the observed Ly$\alpha$ line shape: scattering off HI
outflows is expected to result in asymmetric emission lines (see
Fig~\ref{fig:out}), while peculiar velocities may leave an
intrinsically symmetric emission line symmetric \citep{Cen05}. We
caution that symmetric Ly$\alpha$ emission lines may be observed as a
result of scattering through a static ISM (see Fig~\ref{fig:static}),
but point out that in the latter case the Ly$\alpha$ peak flux density
is redshifted by $\sim 500$ km s$^{-1}$ relative to the galaxy's
systemic redshift. This shift is much larger than that expected for
the model that uses peculiar velocities, which is $\sim 35-70$ km
s$^{-1}$.

\subsection{Comments on the Model}
\label{sec:model}

The model used to generate Ly$\alpha$ line profiles in this paper assumes
a single, thin, compact, spherical shell of HI gas surrounding the
Ly$\alpha$ source \citep[following][]{Ahn03,V06,V08,V10}. Recently,
\citet[][]{Steidel10} described an alternative simple outflow model
that folds in constraints from observed profiles of interstellar metal
absorption lines. This model can also reproduce observed Ly$\alpha$
line shapes in LBGs. In this model, the outflow can extend out to
$\sim 100$ kpc, and outflow velocity increases with radius. It is not
clear whether the models are incompatible: one can imagine
the outflow extends over a wide range of velocities and spatial
scales, but that the outflow's opacity to Ly$\alpha$ photons is
dominated by gas closer to the galaxy that is restricted to a narrower range
of velocities. As a result, it is possible to model the impact of
the outflow on the Ly$\alpha$ radiation field as a single shell that
has a low velocity dispersion. This is clearly an issue that needs
further investigation. However, a more detailed investigation is beyond the
scope of the present paper, especially because our main results are
not likely sensitive to the precise model that one uses to describe
the impact of the outflow on the Ly$\alpha$ radiation field.

In detail, the directly transmitted fraction of Ly$\alpha$ photons
through the IGM at the highest redshifts (the `first' galaxies at
$z>10$) probably depends quantitatively on the precise outflow model
that one uses, since different models may imply different predicted
redshift evolution. For example, in the models of \citet{V06,V08} and
\citet{V10} the emission in the red wing of the line (at $\Delta v
\gsim 500$ km s$^{-1}$) comes from Ly$\alpha$ photons that scatter
back and forth repeatedly between the expanding HI outflow. In the
model of \citet{Steidel10} this frequency `diffusion' does not occur,
and the emission in the red wing of the line reflects the
maximum velocity in the outflow. In this latter model, the flux in the
red wing of the line is suppressed at $z>10$ if the outflow velocities
are much lower at high redshift (see the end of
\S~\ref{sec:outflow}). On the other hand, it is difficult to
completely suppress frequency diffusion. Significant frequency
diffusion occurs even in clumpy outflows with a covering factor of
only $f_{\rm cov} \sim 0.5$ \citep[][their Fig~19]{Hansen06}, which is
less than the outflow covering factor inferred by \citet{Steidel10}
over a significant range of radii. We therefore expect that photons
likely emerge at $\Delta v \gsim 500$ km s$^{-1}$ from the `first'
galaxies in both models, but that the fraction of the flux that
emerges at these frequencies--and therefore $f_{\rm trans}$--depends
on the detailed properties of the HI in outflows in galaxies at these
redshifts. This underlines one of our conclusions presented in
\S~\ref{sec:conc} (also see \S~\ref{sec:EoR}), that it is crucial to
understand the gas kinematics in the ISM (and/or circumgalactic
medium) of high-redshift galaxies in order to infer the properties of
the IGM from the observed Ly$\alpha$ flux.

Our main results regarding galaxies during the later stages of--and
after--reionization ($z\lsim 8$) require less (or no) extrapolation to
higher redshifts, and practically do not depend at all on the precise
underlying model of the Ly$\alpha$ line shape. Instead these results
rely only on the fact that the {\it observed} Ly$\alpha$ line shape
can extend to frequencies that lie well beyond $\Delta v\sim 500$ km
s$^{-1}$ out to $z=5.5$ \citep{V10}.

It has been argued that radiation pressure on dust grains--which in
turn are coupled to the interstellar gas-- provide an important source
of pressure in the ISM of galaxies, and that the observed outflows in
galaxies are driven predominantly by radiation pressure (see
e.g. Murray et al. 2005). Given that the highest redshift galaxies
likely contained little dust (see \S~\ref{sec:dust}), radiation
pressure may be less important in these galaxies, and outflows could
be weakened significantly. Interestingly, especially in these first
galaxies, a significant fraction ($\sim 20\%$) of the total bolometric
luminosity of the galaxy is Ly$\alpha$ line radiation. If this
radiation is `trapped' by a large column of \ion{H}{I} gas, then the
radiation pressure exerted by this Ly$\alpha$ radiation itself becomes
important in driving the \ion{H}{I} gas out \citep{DL08press}, thereby
enhancing the detectability of the Ly$\alpha$ emission.

\section{Conclusions}
\label{sec:conc}
\begin{figure*}
\vbox{\centerline{\epsfig{file=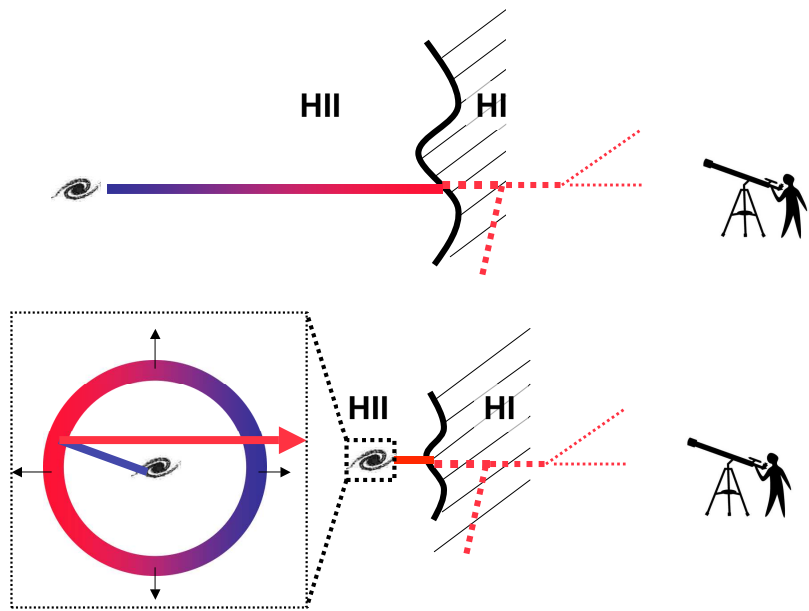,angle=90,width=14.0truecm}}}
\caption[]{Schematic explanation for why outflows promote the
detectability of Ly$\alpha$ emission from galaxies surrounded by
significant amounts of neutral intergalactic gas: In the {\it top
panel} a galaxy is surrounded by a large bubble of ionized gas, which
in turn is surrounded by neutral intergalactic gas. Ly$\alpha$
emission from this galaxy redshifts away from resonance as it
propagates freely through the \ion{H}{II} bubble (as indicated by the
line color). Once the Ly$\alpha$ photons reach the neutral IGM, they
have redshifted far from resonance where the Gunn-Peterson optical
depth is reduced tremendously (see Eq~\ref{eq:redgp}). Because of the
reduced GP optical depth, some fraction of the emitted Ly$\alpha$ is
transmitted to the observer without scattering in the IGM. In this
drawing, the thickness of the line represents the specific intensity
of the Ly$\alpha$ radiation field. The {\it bottom panel} shows that
outflows surrounding star forming regions (represented by the
expanding ring. The far side is receding form the observer and has a
larger redshift, which is represented by the color) can Doppler boost
Ly$\alpha$ photons to frequencies redward of the Ly$\alpha$
resonance. In this scenario, a fraction of Ly$\alpha$ can propagate
directly to the observer {\it without the \ion{H}{II} bubble}.}
\label{fig:scheme}
\end{figure*}
The next generation of telescopes aim to directly observe the first
generation of galaxies that initiated the reionization of our
Universe. The Ly$\alpha$ emission line is robustly predicted to be the
most prominent intrinsic spectral feature of these galaxies. In this
paper we investigated the prospects for detecting this Ly$\alpha$
emission, taking account of radiative transfer effects that are likely
to occur in the interstellar medium (ISM) of these galaxies.

Observed interstellar metal absorption lines (\ion{Si}{II},
\ion{O}{I}, \ion{C}{II}, \ion{Fe}{II} and \ion{Al}{II}) in Lyman Break
Galaxies (LBGs) are typically strongly redshifted relative to the
galaxies' systemic velocity (with a median off-set of $\sim 160$ km
s$^{-1}$), while the Ly$\alpha$ emission line is strongly redshifted
\citep[with a median velocity offset of $\sim 450$ km s$^{-1}$ ][also
see Shapley et al. 2003]{Steidel10}. This suggest that large scale
outflows are ubiquitous in LBGs
\citep{Shapley03,Steidel10}. Furthermore, scattering of Ly$\alpha$
photons by \ion{H}{I} in outflows has successfully explained the
observed Ly$\alpha$ line shapes in Ly$\alpha$ emitting galaxies at
$z=3-6$ \citep[e.g.][]{V06,V08,V10}. Scattering off outflows of
interstellar \ion{H}{I} gas can shift Ly$\alpha$ photons to the red
side of the line before it reaches the IGM (Fig~\ref{fig:out}). At
these frequencies the Gunn-Peterson optical depth may be reduced to
order unity as a result. 

In this paper we investigated the detectability of Ly$\alpha$
radiation under the assumption that the outflows observed at low
redshift also occur in the highest redshift galaxies. We found that
outflows may cause as much as $\gsim 5\%$ of the emitted Ly$\alpha$
radiation to be transmitted directly to the observer, even through a
fully neutral IGM (\S~\ref{sec:outflow}). Since the intrinsic
(restframe) equivalent width of the Ly$\alpha$ line can be as high as
EW$_{\rm int}=1500$ \AA for the first generation of galaxies, the
observed EW$=f_{\rm trans}$EW$_{\rm int}=75(f_{\rm trans}/0.05)({\rm
EW}_{\rm int}/1500\hs{\rm \AA})$ \AA. For comparison, only $4\%$ of
the $z=3$ LBG population have larger EWs \citep{Shapley03}. We showed
that for $f_{\rm trans} \gsim 3\%$ it may be easier to detect
Ly$\alpha$ line emission with NIRSPEC on the {\it James Webb Space
Telescope} (JWST), than continuum radiation with JWST's NIRCAM in the
same integration time. We also note that the next generation of
ground-based 30-m telescopes with diffraction limited AO are expected
to be more (less) sensitive than JWST at $\lambda \gsim 1.3 \mu$m for
$R\gg 100$ ($R \lsim 100$) spectroscopy (see Mountain et al. 2009,
their Fig~2). That is, ground based high-resolution spectroscopic
searches for high redshift galaxies can detect fainter galaxies at a
fixed integration time when $f_{\rm trans} \gsim 5\%$. Such searches
can therefore be competitive with searches that employ the drop-out
technique. Irrespective of the survey strategy that is used to search
for the highest redshift galaxies, the prospect that Ly$\alpha$ can
provide galaxies with spectroscopic redshifts is promising, and
important as no robust predictions exists for the detectability of
other emission lines (\S~\ref{sec:lines}).

This paper has focused on the first generation of galaxies that were
surrounded by a neutral IGM, but our work also applies more broadly.
For example we argued in \S~\ref{sec:EoR} that \ion{H}{I} outflows
promote the detectability of the Ly$\alpha$ emission line during later
stages of reionization when much of the absorption is resonant
absorption in a highly ionized HII region. As a result outflows can
reduce the minimum \ion{H}{II} bubble size that is required to render
LAEs `visible'. This is illustrated schematically in
Figure~\ref{fig:scheme}. Similarly, \ion{H}{I} outflows also promote
the detectability of the Ly$\alpha$ emission line after reionization
has been completed\footnote{Recently, \citet{Z10b} showed that
scattering in the `local' IGM immediately surrounding Ly$\alpha$
sources can introduce a unique anisotropy in the two-point correlation
function of LAEs at $z=5.7$. \citet{Z10b} argued that this
scattering--induced signature is reduced when the intrinsic
(i.e. prior to scattering) Ly$\alpha$ line width is enhanced. Winds
are therefore also expected to reduce this clustering signature. That
is, the clustering of LAEs post-reionization can provide constraints
on the importance of winds in shaping the Ly$\alpha$ line shape.}
(\S~\ref{sec:EoR}).

In summary, radiative transfer effects in the ISM of high redshift
galaxies have been shown to broaden, and--in the case of
outflows--redshift the emergent Ly$\alpha$ flux to a level that allows
$5\%$ or more of the photons to escape absorption by the neutral
IGM. Coupled with the large intrinsic Ly$\alpha$ line EW of the
first generation of galaxies, we have shown that searches for
galaxies in their redshifted Ly$\alpha$ emission line can be
competitive with the drop-out technique out to the highest redshifts
that can be probed observationally in the JWST era.

{ \bf Acknowledgments} MD thanks the School of Physics at the
University of Melbourne, where most of this work was done, for their
kind hospitality. MD is supported by Harvard University funds. We
thank Zolt\'{a}n Haiman \& Zheng Zheng for helpful comments on earlier
versions of this paper.

%{\small

%\appendix
\appendix
\onecolumn
\section{Loeb-Rybicki halos with Outflows}
\label{app:halo}

In the main body of this paper we focused on Ly$\alpha$ radiation that was
transmitted directly to the observer through a fully neutral IGM. This
radiation is confined to an angular region that is set by the physical
scale of the outflow (throughout the paper we referred to this as the
`point source'), and is much more easily detectable than the scattered
Ly$\alpha$ radiation that is in a diffuse halo. For completeness we
show how the surface brightness profile of the scattered radiation is
modified because of the outflow in the galaxy.

\begin{figure*}
\vbox{\centerline{\epsfig{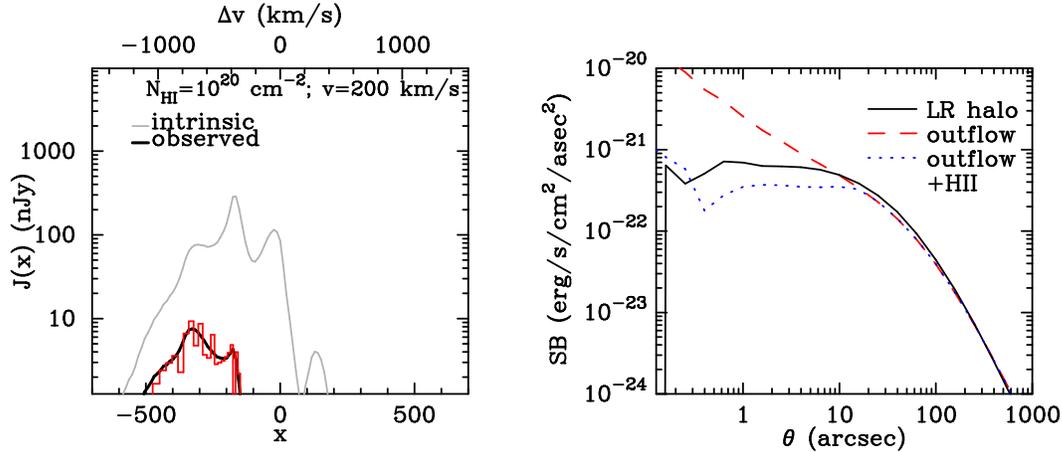}}}
\caption[]{{\it Left panel}: same as Figure~\ref{fig:out}. The {\it
red histogram} shows the spectrum of the `point source` (unscattered
radiation) as extracted from the Monte-Carlo code. {\it Right panel}:
the {\it solid black line} shows the surface brightness profile of the
standard Loeb-Rybicki halo (scattered radiation, as in
Fig~\ref{fig:rl}), while the {\it red dashed line} shows the surface
brightness profile of the scattered radiation for the outflow
model. In the outflow model, the surface brightess profile of the
Ly$\alpha$ halo is boosted significantly at $\theta \lsim 5$
arcsec. However, this boost goes away when a small (radius=50 pkpc) is
present around the galaxy (as shown by the {\it blue dotted
line}). This figure shows that while outflows boost the detectability
of the point source, they do not boost the detectability of the
Ly$\alpha$ LR-halos.}
\label{fig:RLout}
\end{figure*}

In the {\it left panel} of Figure~\ref{fig:RLout} the {\it black
solid line} shows the spectrum of unscattered radiation for the model with
$(N_{\rm HI},v_{\rm sh})=(10^{20}\hs{\rm cm}^{-2},200\hs{\rm km} \hs
{\rm s}^{-1})$ (as in Fig~\ref{fig:out}. The {\it grey solid line} is
again the intrinsic spectrum that emerges from the galaxy). The {\it
red histogram} shows the spectrum of the point source as extracted
from the Monte-Carlo code. The agreement between the Monte-Carlo and
analytic solution (given by Eq~\ref{eq:point}) are again good.

In the {\it right panel} the {\it black solid line} shows the surface
brightness profile of the standard Loeb-Rybicki halo (as in
Fig~\ref{fig:rl}), while the {\it red dashed line} shows the surface
brightness profile of the scattered radiation for the outflow
model. Clearly, the surface brightess profile is boosted significantly
at $\theta \lsim 5$ arcsec. Radiation that escaped from the galaxy far
in the red wing of the line (at $\Delta v\gsim 500$ km s$^{-1}$) is
most likely to scatter in close proximity to the galaxy, where it is
closest to resonance. After this scattering event, these photons again
have a non-negligible probability of propagating directly to the
observer. This boosts the surface brightness in the inner region of
the Ly$\alpha$ halo. We point out that the maximum boost of a factor
of $\sim 10$ is reached at $\theta \lsim 1$ arcsec. In reality we
expect gas to be ionized at such short distances from the
galaxy. Indeed, if we insert a small (radius = 50 pkpc) \ion{H}{II}
bubble around the galaxy, the we find that the boost disappears, and
we almost recover the original surface brightness profile (as
indicated by the {\it blue dotted line}). Larger \ion{H}{II} regions
only suppress the surface brightness profile of the central core. The
main point of this calculation is to demonstrate that the radiative transfer
processes that we invoked to boost the detectability of the `point
source' do not also boost the detectability of the Ly$\alpha$ halos.

\label{lastpage}

\begin{thebibliography}{14}
\expandafter\ifx\csname natexlab\endcsname\relax\def\natexlab#1{#1}\fi
\bibitem[Ahn et al.(2003)]{Ahn03} Ahn, S.-H., Lee, H.-W.,  \& Lee,
H.~M.\ 2003, \mnras, 340, 863

\bibitem[Atek et  al.(2008)]{Atek08} Atek, H., Kunth, D., Hayes, M.,
{\"O}stlin, G., \& Mas-Hesse, J.~M.\ 2008, \aap, 488, 491

\bibitem[Barkana  \& Loeb(2001)]{BL01} Barkana, R., \& Loeb, A.\ 2001,
\physrep, 349, 125

\bibitem[Barkana(2004)]{Infall} Barkana, R.\ 2004, \mnras,  347, 59

\bibitem[Bouwens et al.(2010)]{Bouwens10} Bouwens, R.~J., et al.\
2010, \apjl, 708, L69

\bibitem[Bremer et al.(2004)]{Bremer04} Bremer, M.~N., Jensen,  J.~B.,
Lehnert, M.~D., Schreiber, N.~M.~F.,  \& Douglas, L.\ 2004, \apjl,
615, L1

\bibitem[Brinchmann et al.(2008)]{Br} Brinchmann, J.,  Pettini, M., \&
Charlot, S.\ 2008, \mnras, 385, 769

\bibitem[Bromm et al.(2001)]{Bromm01} Bromm, V., Kudritzki,  R.~P., \&
Loeb, A.\ 2001, \apj, 552, 464

\bibitem[Bromm et al.(2002)]{Bromm02} Bromm, V., Coppi, P.~S.,  \&
Larson, R.~B.\ 2002, \apj, 564, 23

\bibitem[Cen  \& Haiman(2000)]{CH00} Cen, R., \& Haiman, Z.\ 2000,
\apjl, 542, L75

\bibitem[Cen et al.(2005)]{Cen05} Cen, R., Haiman, Z.,  \& Mesinger,
A.\ 2005, \apj, 621, 89

\bibitem[Charlot  \& Fall(1991)]{CF91} Charlot, S., \& Fall, S.~M.\
1991, \apj, 378, 471

\bibitem[Dayal et al.(2010)]{Dayal10} Dayal, P., Maselli, A.,  \&
Ferrara, A.\ 2010, arXiv:1002.0839

\bibitem[Dijkstra et al.(2006)]{D06a} Dijkstra, M., Haiman,  Z., \&
Spaans, M.\ 2006, \apj, 649, 14

\bibitem[Dijkstra \& Wyithe(2006)]{DW06} Dijkstra, M., \& Wyithe, J.S.B.\
2006, \mnras, 372, 1575

\bibitem[Dijkstra et al.(2007a)]{LF} Dijkstra, M., Wyithe,  J.~S.~B.,
\& Haiman, Z.\ 2007a, \mnras, 379, 253

\bibitem[Dijkstra et al.(2007b)]{IGM} Dijkstra, M., Lidz,  A., \&
Wyithe, J.~S.~B.\ 2007b, \mnras, 377, 1175

\bibitem[Dijkstra  \& Loeb(2008)]{DL08press} Dijkstra, M., \& Loeb,
A.\ 2008, \mnras, 391, 457

\bibitem[Furlanetto et al.(2006)]{F06} Furlanetto, S.~R., Zaldarriaga,
M., \& Hernquist, L.\ 2006, \mnras, 365, 1012

\bibitem[Haiman  \& Spaans(1999)]{HS99} Haiman, Z., \& Spaans, M.\
1999, \apj, 518, 138

\bibitem[Haiman(2002)]{Haiman02} Haiman, Z.\ 2002, \apjl, 576,  L1

\bibitem[Hansen  \& Oh(2006)]{Hansen06} Hansen, M., \& Oh, S.~P.\
2006, \mnras, 367, 979

\bibitem[Harrington(1973)]{Harrington73} Harrington, J.~P.\ 1973,
\mnras, 162, 43

\bibitem[Hartmann et al.(1984)]{Hartmann84} Hartmann, L.~W.,  Huchra,
J.~P., \& Geller, M.~J.\ 1984, \apj, 287, 487

\bibitem[Hartmann et al.(1988)]{Hartmann88} Hartmann, L.~W.,  Huchra,
J.~P., Geller, M.~J., O'Brien, P.,  \& Wilson, R.\ 1988, \apj, 326,
101

\bibitem[Hayes et al.(2007)]{2007MNRAS.382.1465H} Hayes, M.,
{\"O}stlin,  G., Atek, H., Kunth, D., Mas-Hesse, J.~M., Leitherer, C.,
Jim{\'e}nez-Bail{\'o}n, E., \& Adamo, A.\ 2007, \mnras, 382, 1465

\bibitem[Hayes et al.(2010)]{2010arXiv1002.4876H} Hayes, M., et al.\
2010,  arXiv:1002.4876

\bibitem[Huchra(1977)]{Huchra77} Huchra, J.~P.\ 1977, \apj, 217, 
928 

\bibitem[Iliev et al.(2008)]{Iliev08} Iliev, I.~T., Shapiro,  P.~R.,
McDonald, P., Mellema, G., \& Pen, U.-L.\ 2008, \mnras, 391, 63

\bibitem[Jimenez  \& Haiman(2006)]{JH06} Jimenez, R., \& Haiman, Z.\
2006, \nat, 440, 501

%\bibitem[Dijkstra 
%\& Loeb(2008b)]{DL08pol} Dijkstra, M., \& Loeb, A.\ 2008b, \mnras, 386, 492 
\bibitem[Johnson(2009)]{Johnson09} Johnson, J.~L.\ 2009,
arXiv:0911.1294


\bibitem[Johnson et al.(2009)]{J09b} Johnson, J.~L., Greif,  T.~H.,
Bromm, V., Klessen, R.~S., \& Ippolito, J.\ 2009, \mnras, 399, 37

\bibitem[Kashikawa et al.(2006)]{Ka06} Kashikawa, N., et  al.\ 2006,
\apj, 648, 7

\bibitem[Kobayashi  \& Kamaya(2004)]{Ko04} Kobayashi, M.~A.~R., \&
Kamaya, H.\ 2004, \apj, 600, 564

\bibitem[Kobayashi et al.(2006)]{Ko06} Kobayashi, M.~A.~R.,  Kamaya,
H., \& Yonehara, A.\ 2006, \apj, 636, 1

\bibitem[Kobayashi et al.(2010)]{Ko10} Kobayashi, M.~A.~R.,  Totani,
T., \& Nagashima, M.\ 2010, \apj, 708, 1119

\bibitem[Komatsu et al.(2009)]{Komatsu08} Komatsu, E., et al.\  2009,
\apjs, 180, 330
\bibitem[Kunth et  al.(1994)]{Kunth94} Kunth, D., Lequeux, J.,
Sargent, W.~L.~W., \& Viallefond, F.\ 1994, \aap, 282, 709



\bibitem[Kunth et  al.(1998)]{Kunth98} Kunth, D., Mas-Hesse, J.~M.,
Terlevich, E., Terlevich, R., Lequeux, J., \& Fall, S.~M.\ 1998, \aap,
334, 11

\bibitem[Larson(1998)]{La98} Larson, R.~B.\ 1998, \mnras,  301, 569

\bibitem[Laursen et al.(2009)]{Laursen09} Laursen, P., 
Sommer-Larsen, J., \& Andersen, A.~C.\ 2009, \apj, 704, 1640 

\bibitem[Lequeux et  al.(1995)]{L95} Lequeux, J., Kunth, D.,
Mas-Hesse, J.~M., \& Sargent, W.~L.~W.\ 1995, \aap, 301, 18

\bibitem[Loeb  \& Rybicki(1999)]{LR99} Loeb, A., \& Rybicki, G.~B.\
1999, \apj, 524, 527
\bibitem[Meier(1976)]{1976ApJ...207..343M} Meier, D.~L.\ 1976, \apj,
207,  343
\bibitem[Malhotra  \& Rhoads(2004)]{MR04} Malhotra, S., \& Rhoads,
J.~E.\ 2004, \apjl, 617, L5

\bibitem[Mao et al.(2007)]{Mao07} Mao, J., Lapi, A., Granato,  G.~L.,
de Zotti, G., \& Danese, L.\ 2007, \apj, 667, 655

\bibitem[Martin(2005)]{Martin05} Martin, C.~L.\ 2005, \apj, 621,  227

\bibitem[McQuinn et al.(2007)]{McQ} McQuinn, M., Hernquist,  L.,
Zaldarriaga, M., \& Dutta, S.\ 2007, \mnras, 381, 75

\bibitem[Meier(1976)]{1976ApJ...207..343M} Meier, D.~L.\ 1976, \apj,
207,  343

\bibitem[Mesinger  \& Furlanetto(2008)]{Mesinger} Mesinger, A., \&
Furlanetto, S.~R.\ 2008, \mnras, 386, 1990

\bibitem[Miralda-Escude(1998)]{M98} Miralda-Escude, J.\  1998, \apj,
501, 15

\bibitem[Mountain et al.(2009)]{2009astro2010T..12M} Mountain, M., et al.\ 
2009, astro2010: The Astronomy and Astrophysics Decadal Survey, 2010, 12 

\bibitem[Murray et al.(2005)]{2005ApJ...618..569M} Murray, N., Quataert, 
E., \& Thompson, T.~A.\ 2005, \apj, 618, 569 

\bibitem[Neufeld(1990)]{Neufeld90} Neufeld, D.~A.\ 1990, \apj,  350,
216
\bibitem[Neufeld(1991)]{Neufeld91} Neufeld, D.~A.\ 1991, \apjl,  370,
L85

\bibitem[Oh et al.(2001)]{Oh01} Oh, S.~P., Haiman, Z.,  \& Rees,
M.~J.\ 2001, \apj, 553, 73

\bibitem[Ota et al.(2008)]{Ota08} Ota, K., et al.\ 2008, 
\apj, 677, 12 

\bibitem[Ouchi et al. (2010)]{Ouchi10} Ouchi, M., et al., to be submitted to MNRAS

\bibitem[Partridge  \& Peebles(1967)]{1967ApJ...147..868P} Partridge,
R.~B., \& Peebles, P.~J.~E.\ 1967, \apj, 147, 868

\bibitem[Pell{\'o} et  al.(2004)]{Pello04} Pell{\'o}, R., Schaerer,
D., Richard, J., Le Borgne, J.-F., \& Kneib, J.-P.\ 2004, \aap, 416,
L35

\bibitem[Pritchard 
\& Loeb(2008)]{2008PhRvD..78j3511P} Pritchard, J.~R., \& Loeb, A.\ 2008, Phys Rev D, 78, 103511 

\bibitem[Rauch et al.(2008)]{Rauch08} Rauch, M., et al.\ 2008,  \apj,
681, 856

\bibitem[Ricotti et al.(2008)]{Ricotti08} Ricotti, M., Gnedin,  N.~Y.,
\& Shull, J.~M.\ 2008, \apj, 685, 21

\bibitem[Rybicki  \& Lightman(1979)]{RL79} Rybicki, G.~B., \&
Lightman, A.~P.\ 1979, New York, Wiley-Interscience, 1979.~393 p.,

\bibitem[Scarlata et al.(2009)]{Scarlata09} Scarlata, C., et al.\
2009, \apjl, 704, L98

\bibitem[Schaerer(2002)]{S02} Schaerer, D.\ 2002, \aap, 382, 28

\bibitem[Schaerer(2003)]{S03} Schaerer, D.\ 2003, \aap,  397, 527
\bibitem[Schaerer(2008)]{S08} Schaerer, D.\ 2008, IAU  Symposium, 255,
66

\bibitem[Shapley et al.(2003)]{Shapley03} Shapley, A.~E.,  Steidel,
C.~C., Pettini, M., \& Adelberger, K.~L.\ 2003, \apj, 588, 65

\bibitem[Shimasaku et al.(2006)]{Shima06} Shimasaku, K., et  al.\
2006, \pasj, 58, 313

\bibitem[Steidel et al.(2010)]{Steidel10} Steidel, C.~C., Erb,  D.~K.,
Shapley, A.~E., Pettini, M., Reddy, N.~A., Bogosavljevi{\'c}, M.,
Rudie, G.~C., \& Rakic, O.\ 2010, arXiv:1003.0679

\bibitem[Tumlinson  \& Shull(2000)]{TS00} Tumlinson, J., \& Shull,
J.~M.\ 2000, \apjl, 528, L65

\bibitem[Vanzella et al.(2009)]{V10} Vanzella, E., et al.\  2009,
arXiv:0912.3007


\bibitem[Verhamme et al.(2006)]{V06} Verhamme, A.,  Schaerer, D., \&
Maselli, A.\ 2006, \aap, 460, 397

\bibitem[Verhamme et  al.(2008)]{V08} Verhamme, A., Schaerer, D.,
Atek, H., \& Tapken, C.\ 2008, \aap, 491, 89
\bibitem[Weatherley et  al.(2004)]{W04} Weatherley, S.~J., Warren,
S.~J., \& Babbedge, T.~S.~R.\ 2004, \aap, 428, L29

\bibitem[Zheng et al.(2010a)]{ZZ} Zheng, Z., Cen, R., Trac,  H., \&
Miralda-Escude, J.\, 2010a, arXiv:0910.2712

\bibitem[Zheng et al.(2010b)]{Z10b} Zheng, Z., Cen, R., Trac, H., \&
Miralda-Escude, J.\ 2010b, submitted to ApJ




\end{thebibliography}
\end{document}